\documentclass[conference]{IEEEtran}
\IEEEoverridecommandlockouts

\usepackage{hyperref}
\hypersetup{%
  unicode=true,%
  pdfstartview={FitH},%
  colorlinks=true,%
  citecolor=black,%
  filecolor=black,%
  linkcolor=black,%
  urlcolor=black%
}

\ifCLASSINFOpdf
   \usepackage[pdftex]{graphicx}
\else
\fi

\usepackage[cmex10]{amsmath}
\usepackage{wrapfig}                    
\usepackage[ruled]{algorithm2e}         
\usepackage{setspace}                   
\usepackage{dcolumn}                    
\usepackage[font=small]{caption,subfig} 
\usepackage{amsfonts}                   
\newcommand{\note}[1]{}                 

\newcommand{\squishlist}{
  \begin{list}{$\bullet$}{
    \setlength{\itemsep}{0pt}
    \setlength{\parsep}{1pt}
    \setlength{\topsep}{1pt}
    \setlength{\partopsep}{0pt}
    \setlength{\leftmargin}{\parindent}
    \setlength{\labelwidth}{1em}
    \setlength{\listparindent}{\parindent}
    \setlength{\labelsep}{0.5em} } }

\newcommand{\squishend}{
  \end{list} }

\newcommand{\ignore}[1]{}

\title{\huge The RowHammer Problem \\ and Other Issues We May Face as Memory Becomes Denser}

\author{
\IEEEauthorblockN{Onur Mutlu}
\IEEEauthorblockA{ETH Z\"{u}rich}
\href{mailto:onur.mutlu@inf.ethz.ch}{onur.mutlu@inf.ethz.ch}\\
\href{https://people.inf.ethz.ch/omutlu}{https://people.inf.ethz.ch/omutlu}
}

\IEEEpeerreviewmaketitle

\begin{document}
\begin{spacing}{0.93}
\maketitle

\thispagestyle{plain}
\pagestyle{plain}
\setcounter{page}{1}
\pagenumbering{arabic}

\small

\begin{abstract}


As memory scales down to smaller technology nodes, new failure
mechanisms emerge that threaten its correct operation. If such failure
mechanisms are not anticipated and corrected, they can not only
degrade system reliability and availability but also, perhaps even
more importantly, open up security vulnerabilities: a malicious
attacker can exploit the exposed failure mechanism to take over the
entire system. As such, new failure mechanisms in memory can become
practical and significant threats to system security.

In this work, we discuss the RowHammer problem in DRAM, which is a
prime (and perhaps the first) example of how a circuit-level failure
mechanism in DRAM can cause a practical and widespread system security
vulnerability. RowHammer, as it is popularly referred to, is the
phenomenon that repeatedly accessing a row in a modern DRAM chip
causes bit flips in physically-adjacent rows at consistently
predictable bit locations. It is caused by a hardware failure
mechanism called DRAM disturbance errors, which is a manifestation of
circuit-level cell-to-cell interference in a scaled memory
technology. Researchers from Google Project Zero recently demonstrated
that this hardware failure mechanism can be effectively exploited by
user-level programs to gain kernel privileges on real systems. Several
other recent works demonstrated other practical attacks exploiting
RowHammer. These include remote takeover of a server vulnerable to
RowHammer, takeover of a victim virtual machine by another virtual
machine running on the same system, and takeover of a mobile device by
a malicious user-level application that requires no permissions.

We analyze the root causes of the RowHammer problem and examine
various solutions. We also discuss what other vulnerabilities may be
lurking in DRAM and other types of memories, e.g., NAND flash memory
or Phase Change Memory, that can potentially threaten the foundations
of secure systems, as the memory technologies scale to higher
densities. We conclude by describing and advocating a principled
approach to memory reliability and security research that can enable
us to better anticipate and prevent such vulnerabilities.

\end{abstract}


\section{Introduction}

Memory is a key component of all modern computing systems, often
determining the overall performance, energy efficiency, and
reliability characteristics of the entire system. The push for
increasing the density of modern memory technologies via technology
scaling, which has resulted in higher capacity (i.e., density) memory
and storage at lower cost, has enabled large leaps in the performance
of modern computers~\cite{mutlu-imw13}. This positive trend is clearly
visible in especially the dominant main memory and solid-state storage
technologies of today, i.e.,
DRAM~\cite{tldram,kevinchang-sigmetrics16} and NAND flash
memory~\cite{cai-date12}, respectively. Unfortunately, the same push
has also greatly decreased the reliability of modern memory
technologies, due to the increasingly smaller memory cell size and
increasingly smaller amount of charge that is maintainable in the
cell, which makes the memory cell much more vulnerable to various
failure mechanisms and noise and interference sources, both in
DRAM~\cite{dram-isca2013,rowhammer-isca2014,samira-sigmetrics14,kang-memforum2014}
and NAND
flash~\cite{cai-date12,cai-date13,cai-iccd13,cai-hpca15,cai-hpca17,
  cai-iccd12, cai-itj2013, cai-dsn15, cai-sigmetrics14, yixin-jsac16}.

In this work, and the associated invited special session talk, we
discuss the effects of reduced memory reliability on system
security. As memory scales down to smaller technology nodes, new
failure mechanisms emerge that threaten its correct operation. If such
failure mechanisms are not anticipated and corrected, they can not
only degrade system reliability and availability, but also, perhaps
even more importantly, open up security vulnerabilities: a malicious
attacker can exploit the exposed failure mechanism to take over the
entire system. As such, new failure mechanisms in memory can become
practical and significant threats to system security.

We first discuss the RowHammer problem in DRAM, as a prime example of
such a failure mechanism. We believe RowHammer is the first
demonstration of how a circuit-level failure mechanism in DRAM can
cause a practical and widespread system security vulnerability
(Section~\ref{sec:rowhammer-problem}). After analyzing RowHammer in
detail, we describe solutions to it
(Section~\ref{sec:rowhammer-solutions}). We then turn our attention to
other vulnerabilities that may be present or become present in DRAM
and other types of memories (Section~\ref{sec:other-problems}), e.g.,
NAND flash memory or Phase Change Memory, that can potentially
threaten the foundations of secure systems, as the memory technologies
scale to higher densities. We conclude by describing and advocating a
principled approach to memory reliability and security research that
can enable us to better anticipate and prevent such vulnerabilities
(Section~\ref{sec:prevention}).

\section{The RowHammer Problem}
\label{sec:rowhammer-problem}

\begin{figure}[t]
\centering
\includegraphics{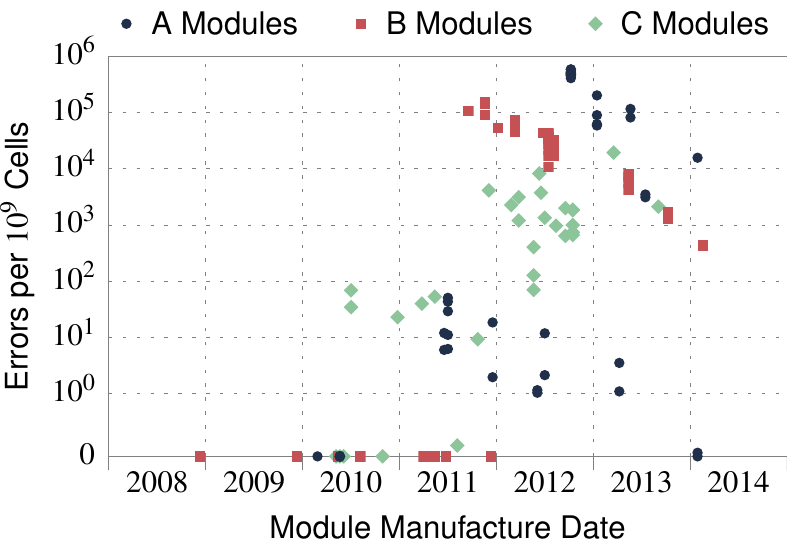}
\caption{RowHammer error rate vs.~manufacturing dates of 129 DRAM modules we tested (reproduced from~\cite{rowhammer-isca2014}).}
\label{fig:errors_vs_date}
\end{figure}

Memory isolation is a key property of a reliable and secure computing
system. An access to one memory address should not have unintended
side effects on data stored in other addresses. However, as process
technology scales down to smaller dimensions, memory chips become more
vulnerable to {\em disturbance}, a phenomenon in which different
memory cells interfere with each others' operation. We have shown, in
our ISCA 2014 paper~\cite{rowhammer-isca2014}, the existence of {\em
  disturbance errors} in commodity DRAM chips that are sold and used
in the field today. Repeatedly reading from the same address in DRAM
could corrupt data in nearby addresses. Specifically, when a DRAM row
is opened (i.e., activated) and closed (i.e., precharged) repeatedly
(i.e., {\em hammered}), enough times within a DRAM refresh interval,
one or more bits in physically-adjacent DRAM rows can be flipped to
the wrong value. This DRAM failure mode is now popularly called {\em
  RowHammer}~\cite{rowhammer-arxiv16,rowhammer-wikipedia,rh-discuss,rh-twitter,rowhammer-thirdio,anvil,rowhammer-js,google-project-zero,google-rh-blackhat,dedup-est-machina,flip-feng-shui,drammer}. Using
an FPGA-based experimental DRAM testing infrastructure, which we
originally developed for testing retention time issues in
DRAM~\cite{dram-isca2013},\footnote{This infrastructure is currently
  released to the public, and is described in detail in our HPCA 2017
  paper~\cite{softmc}. The infrastructure has enabled many
  studies~\cite{aldram, dram-isca2013, rowhammer-isca2014, softmc,
    kevinchang-sigmetrics16, memcon-cal16, samira-sigmetrics14,
    avatar-dsn15, khan-dsn16} into the failure and performance
  characteristics of modern DRAM, which were previously not well
  understood.} we tested 129 DRAM modules manufactured by three major
manufacturers (A, B, C) in seven recent years (2008--2014) and found
that 110 of them exhibited RowHammer errors, the earliest of which
dates back to 2010. This is illustrated in
Figure~\ref{fig:errors_vs_date}, which shows the error rates we found
in all 129 modules we tested where modules are categorized based on
manufacturing date.\footnote{Test details and experimental setup,
  along with a listing of all modules and their characteristics, are
  reported in our original RowHammer paper~\cite{rowhammer-isca2014}.}
In particular, {\em all} DRAM modules from 2012--2013 were vulnerable
to RowHammer, indicating that RowHammer is a recent phenomenon
affecting more advanced process technology generations.

\subsection{User-Level RowHammer}

We have also demonstrated that a very simple user-level
program~\cite{rowhammer-isca2014,safari-rowhammer} can reliably and
consistently induce RowHammer errors in three commodity AMD and Intel
systems using vulnerable DRAM modules. We released the source code of
this program~\cite{safari-rowhammer}, which Google Project Zero later
enhanced~\cite{google-rowhammer-test}. Using our user-level RowHammer
test program, we showed that RowHammer errors violate two invariants
that memory should provide: (i) a read access should not modify data
at any address and (ii) a write access should modify data only at the
address that it is supposed to write to. As long as a row is
repeatedly opened, both read and write accesses can induce RowHammer
errors, all of which occur in rows other than the one that is being
accessed. Since different DRAM rows are mapped (by the memory
controller) to different software pages, our user-level program could
reliably corrupt specific bits in pages belonging to other
programs. As a result, RowHammer errors can be exploited by a
malicious program to breach memory protection and compromise the
system. In fact, we hypothesized, in our ISCA 2014 paper, that our
user-level program, with some engineering effort, could be developed
into a {\em disturbance attack} that injects errors into other
programs, crashes the system, or hijacks control of the system. We
left such research for the future since our primary objective was to
understand and prevent RowHammer
errors~\cite{rowhammer-isca2014}.\footnote{Our ISCA 2014
  paper~\cite{rowhammer-isca2014} provides a detailed analysis of
  various characteristics of RowHammer, including its data pattern
  dependence, relationship with leaky cells, repeatability of errors,
  and various circuit-level causes of the phenomenon. We omit these
  analyses here and focus on security vulnerabilities and prevention
  of RowHammer, and refer the reader to~\cite{rowhammer-isca2014} for
  a rigorous treatment of the characteristics and causes of the
  RowHammer phenomenon.}

\subsection{RowHammer as a Security Threat}

RowHammer exposes a {\em security threat} since it leads to a breach
of memory isolation, where accesses to one row (e.g., an OS page)
modifies the data stored in another memory row (e.g., another OS
page). As indicated above, malicious software can be written to take
advantage of these disturbance errors. We call these {\em disturbance
  attacks}~\cite{rowhammer-isca2014}, or {\em RowHammer attacks}. Such
attacks can be used to corrupt system memory, crash a system, or take
over the entire system. Confirming the predictions of our ISCA
paper~\cite{rowhammer-isca2014}, researchers from Google Project Zero
developed a user-level attack that exploits RowHammer to gain kernel
privileges and thus take over an entire
system~\cite{google-project-zero, google-rh-blackhat}. More recently,
researchers showed that RowHammer can be exploited in various ways to
take over various classes of systems. For example, RowHammer
vulnerability in a remote server can be exploited to remotely take
over the server via the use of JavaScript~\cite{rowhammer-js}. Or, a
virtual machine can take over a victim virtual machine running on the
same system by inducing RowHammer errors in the victim virtual
machine's memory space~\cite{flip-feng-shui}. Or, a malicious
application that requires no permissions can take control of a mobile
device by exploiting RowHammer, as demonstrated in real Android
devices~\cite{drammer}. Or, an attacker can gain arbitrary read and
write access in a web browser by exploiting RowHammer together with
memory deduplication, as demonstrated on a real Microsoft Windows 10
system~\cite{dedup-est-machina}. As such, the RowHammer problem has
widespread and profound real implications on system security,
threatening the foundations of memory isolation on top of which modern
system security principles are built.

As described above, RowHammer has recently been the subject of many
popular analyses and discussions on hardware-induced security
problems~\cite{rowhammer-arxiv16,rowhammer-wikipedia,rh-discuss,rh-twitter,rh-zdnet1,rowhammer-thirdio,anvil,rowhammer-js,drammer,flip-feng-shui,dedup-est-machina,google-rh-blackhat}
as well as the prime vulnerability exploited by various software-level
attacks that rely on no permissions or software
vulnerabilities~\cite{rowhammer-js,drammer,flip-feng-shui,dedup-est-machina}. As
a result of the severity of the security problem, several major system
manufacturers increased DRAM refresh rates to reduce the probability
of occurrence of
RowHammer~\cite{rh-apple,rh-hp,rh-lenovo,rh-cisco}. And, multiple
memory test programs have been augmented to test for RowHammer
errors~\cite{rh-passmark,rh-futureplus,drammer}. Unfortunately, some
recent reports suggest that even state-of-the-art DDR4 DRAM chips are
vulnerable to RowHammer~\cite{rowhammer-thirdio}, which suggests that
the security vulnerabilities might continue in the current generation
of DRAM chips as well. As such, it is critical to investigate
solutions to the RowHammer vulnerability.

\subsection{Solutions to RowHammer}
\label{sec:rowhammer-solutions}
           
Given that it is such a critical vulnerability, it is important to
find both {\em immediate} and {\em long-term} solutions to the
RowHammer problem (as well as related problems that might cause
similar vulnerabilities). The goal of the immediate solutions is to
ensure that existing systems are patched such that the vulnerable DRAM
devices that are already in the field cannot be exploited. The goal of
the long-term solutions is to ensure that future DRAM devices do not
suffer from the RowHammer problem when they are released into the
field.

Given that immediate solutions require mechanisms that already exist
in systems operating in the field, they are fundamentally more
limited. A popular immediate solution, described and analyzed by our
ISCA 2014 paper~\cite{rowhammer-isca2014}, is to increase the refresh
rate of memory such that the probability of inducing a RowHammer error
before DRAM cells get refreshed is reduced. Several major system
manufacturers have adopted this solution and released security patches
that increased DRAM refresh rates
(e.g.,~\cite{rh-apple,rh-hp,rh-lenovo,rh-cisco}) in the memory
controllers. While this solution might be practical and effective in
reducing the vulnerability, it has the significant drawbacks of
increasing energy/power consumption, reducing system performance, and
degrading quality of service experienced by user programs. Our paper
shows that the refresh rate needs to be increased by 7X if we want to
eliminate all RowHammer-induced errors we saw in our tests of 129 DRAM
modules. Since DRAM refresh is already a significant
burden~\cite{raidr,darp-hpca2014,kang-memforum2014,samira-sigmetrics14,avatar-dsn15}
on energy consumption, performance, and quality of service, increasing
it by any significant amount would only exacerbate the problem. Yet,
increased refresh rate is likely the most practical {\em immediate}
solution to RowHammer.

Other immediate solutions modify the software. For example, ANVIL
proposes software-based detection of RowHammer attacks by monitoring
via hardware performance counters and selective explicit refreshing of
victim rows that are found to be
vulnerable~\cite{anvil}. Unfortunately, such solutions require
modifications to system software and might be intrusive to system
operation (yet they are a promising area of research).

Our ISCA 2014 paper~\cite{rowhammer-isca2014} discusses and analyzes
seven long-term countermeasures to the RowHammer problem.  The first
six solutions are: 1) making better DRAM chips that are not
vulnerable, 2) using error correcting codes (ECC), 3) increasing the
refresh rate (as discussed above), 4) remapping RowHammer-prone cells
after manufacturing, 5) remapping/retiring RowHammer-prone cells at
the user level during operation, 6) accurately identifying hammered
rows during runtime and refreshing their neighbors. None of these
first six solutions are very desirable as they come at significant
power, performance or cost overheads. We already discussed the
overheads of increasing the refresh rates across the board. Similarly,
simple SECDED ECC (an example of the second solution above), as
employed in many systems, is not enough to prevent all RowHammer
errors, as some cache blocks experience two or more bit flips, which
are not correctable by SECDED ECC, as we have
shown~\cite{rowhammer-isca2014}. Thus, stronger ECC is likely required
to correct RowHammer errors, which comes at the cost of additional
energy, performance, cost, and DRAM capacity overheads. Alternatively,
the sixth solution described above, i.e., accurately identifying a row
as a hammered row requires keeping track of access counters for a
large number of rows in the memory controller~\cite{moin-rowhammer},
leading to very large hardware area and power consumption, and
potentially performance, overheads.

We believe the long-term solution to RowHammer can actually be very
simple and low cost: when the memory controller closes a row
(after it was activated), it, with a very low probability, refreshes
the adjacent rows. The probability value is a parameter determined by
the system designer or provided programmatically, if needed, to trade
off between performance overhead and vulnerability protection
guarantees. We show that this probabilistic solution, called {\em PARA
  (Probabilistic Adjacent Row Activation)}, is extremely effective: it
eliminates the RowHammer vulnerability, providing much higher
reliability guarantees than modern hard disks today, while requiring
no storage cost and having negligible performance and energy
overheads~\cite{rowhammer-isca2014}.

PARA is not immediately implementable because it requires changes to
either the memory controllers or the DRAM chips, depending on where it
is implemented. If PARA is implemented in the memory controller, the
memory controller needs to obtain information on which rows are
adjacent to each other in a DRAM bank. This information is currently
{\em unknown} to the memory controller as DRAM manufacturers can
internally remap rows to other
locations~\cite{dram-isca2013,rowhammer-isca2014,khan-dsn16,memcon-cal16,divadram-arxiv16}
for various reasons, including for tolerating various types of
faults. However, this information can be simply provided by the DRAM
chip to the memory controller using the serial presence detect (SPD)
read-only memory present in modern DRAM modules, as described in our
ISCA 2014 paper~\cite{rowhammer-isca2014}. If PARA is implemented in
the DRAM chip, then the hardware interface to the DRAM chip should be
such that it allows DRAM-internal refresh operations that are not
initiated by an external memory controller. This could be achieved
with the addition of a new DRAM command, like the {\em targeted
  refresh} command proposed in a patent by
Intel~\cite{intel-rh-refresh}. In 3D-stacked memory
technologies~\cite{ramulator,smla-taco16}, e.g., HBM (High Bandwidth
Memory)~\cite{hbm,smla-taco16} or HMC (Hybrid Memory Cube)~\cite{hmc},
which combine logic and memory in a tightly integrated fashion, the
logic layer can be easily modified to implement PARA.

All these implementations of the promising PARA solution are examples
of much better cooperation between memory controller and the DRAM
chips.  Regardless of the exact implementation, we believe RowHammer,
and other upcoming reliability vulnerabilities like RowHammer, can be
much more easily found, mitigated, and prevented with better
cooperation between and co-design of system and memory, i.e., {\em
  system-memory co-design}~\cite{mutlu-imw13}. System-memory co-design
is explored by recent works for mitigating various DRAM scaling
issues, including retention failures and performance
problems~\cite{raidr, salp, mutlu-imw13, kang-memforum2014,
  superfri14, hrm-dsn2014, samira-sigmetrics14, avatar-dsn15,
  khan-dsn16, memcon-cal16, aldram, tldram, rowclone, lisa,
  kevinchang-sigmetrics16, dram-isca2013, darp-hpca2014,
  rowhammer-isca2014, divadram-arxiv16, chargecache-hpca16, gs-dram,
  ddma-pact15, vivek-and-or}. Taking the system-memory co-design
approach further, providing more intelligence and
configurability/programmability in the memory controller can greatly
ease the tolerance to errors like RowHammer: when a new failure
mechanism in memory is discovered, the memory controller can be
configured/programmed/patched to execute specialized functions to
profile and correct for such mechanisms. We believe this direction is
very promising, and several works have explored {\em online profiling}
mechanisms for fixing retention
errors~\cite{samira-sigmetrics14,avatar-dsn15,khan-dsn16,
  memcon-cal16} and reducing latency~\cite{divadram-arxiv16}. These
works provide examples of how an intelligent memory controller can
alleviate the retention failures, and thus the DRAM refresh
problem~\cite{raidr, dram-isca2013}, as well as the DRAM latency
problem~\cite{tldram, aldram}.

\subsection{Putting RowHammer into Context}

Springing off from the stir created by RowHammer, we take a step back
and argue that there is little that is surprising about the fact that
we are seeing disturbance errors in the heavily-scaled DRAM chips of
today. Disturbance errors are a general class of reliability problems
that is present in not only DRAM, but also other memory and storage
technologies. All scaled memory technologies, including
SRAM~\cite{chen05, guo09, kim11}, flash~\cite{cai-date12, cai-date13,
  cai-iccd13, cai-dsn15, cooke07}, and hard disk drives~\cite{jiang03,
  tang08, wood09}, exhibit such disturbance problems. In fact, our
recent work at DSN 2015~\cite{cai-dsn15} experimentally characterizes
the read disturb errors in flash memory, shows that the problem is
widespread in flash memory chips, and develops mechanisms to correct
such errors in the flash memory controller. Even though the mechanisms
that cause the bit flips are different in different technologies, the
high-level root cause of the problem, {\em cell-to-cell interference},
i.e., that the memory cells are too close to each other, is a
fundamental issue that appears and will appear in any technology that
scales down to small enough technology nodes. Thus, we should expect
such problems to continue as we scale any memory technology, including
emerging ones, to higher densities.

What sets DRAM disturbance errors apart from other technologies'
disturbance errors is that in modern DRAM, as opposed to other
technologies, error correction mechanisms are not commonly employed
(either in the memory controller or the memory chip). The success of
DRAM scaling until recently has {\em not} relied on a memory
controller that corrects errors (other than performing periodic
refresh). Instead, DRAM chips were implicitly assumed to be error-free
and did {\em not} require the help of the controller to operate
correctly. Thus, such errors were perhaps not as easily anticipated
and corrected within the context of DRAM. In contrast, the success of
other technologies, e.g., flash memory and hard disks, has heavily
relied on the existence of an intelligent controller that plays a key
role in correcting errors and making up for reliability problems of
the memory chips themselves. This has not only enabled the correct
operation of assumed-faulty memory chips but also enabled a mindset
where the controllers are co-designed with the chips themselves,
covering up the memory technology's deficiencies and hence perhaps
enabling better anticipation of errors with technology scaling. This
approach is very prominent in modern SSDs (solid state drives), for
example, where the flash memory controller employs a wide variety of
error mitigation and correction mechanisms~\cite{cai-iccd12,
  cai-date12, cai-date13, cai-iccd13, cai-sigmetrics14, cai-dsn15,
  cai-hpca17, cai-hpca15, cai-itj2013, yixin-jsac16}, including not
only sophisticated ECC mechanisms but also targeted voltage
optimization, retention mitigation and disturbance mitigation
techniques. We believe changing the mindset in modern DRAM to a
similar mindset of {\em assumed-faulty memory chip and an intelligent
  memory controller that makes it operate correctly} can not only
enable better anticipation and correction of future issues like
RowHammer but also better scaling of DRAM into future technology
nodes~\cite{mutlu-imw13}.


\section{Other Potential Vulnerabilities}
\label{sec:other-problems}

We believe that, as memory technologies scale to higher densities,
other problems may start appearing (or may already be going unnoticed)
that can potentially threaten the foundations of secure systems. There
have been recent large-scale field studies of memory errors showing
that both DRAM and NAND flash memory technologies are becoming less
reliable~\cite{justin-memerrors-dsn15,dram-field-analysis2,
  dram-field-analysis3, dram-field-analysis4,
  justin-flash-sigmetrics15, flash-field-analysis2}. As detailed
experimental analyses of real DRAM and NAND flash chips show, both
technologies are becoming much more vulnerable to cell-to-cell
interference effects~\cite{rowhammer-isca2014, cai-dsn15,
  cai-sigmetrics14, cai-iccd13, cai-date12,cai-date13, flash-fms-talk,
  yixin-jsac16, cai-hpca17}, data retention is becoming significantly
more difficult in both technologies~\cite{raidr,samira-sigmetrics14,
  dram-isca2013, khan-dsn16, avatar-dsn15, darp-hpca2014,
  kang-memforum2014, mandelman-jrd02, cai-hpca15, cai-iccd12,
  warm-msst15, cai-date12,cai-date13,cai-itj2013, flash-fms-talk,
  memcon-cal16}, and error variation within and across chips is
increasingly
prominent~\cite{dram-isca2013,aldram,kevinchang-sigmetrics16,dram-process-variation-3,cai-date12,cai-date13,divadram-arxiv16}. Emerging
memory technologies~\cite{mutlu-imw13,meza-weed13}, such as
Phase-Change
Memory~\cite{pcm-isca09,zhou-isca09,moin-isca09,moin-micro09,wong-pcm,raoux-pcm,pcm-ieeemicro10,pcm-cacm10,
  justin-taco14, rbla},
STT-MRAM~\cite{chen-ieeetmag10,kultursay-ispass13}, and
RRAM/ReRAM/memristors~\cite{wong-rram} are likely to exhibit similar
and perhaps even more exacerbated reliability issues. We believe, if
not carefully accounted for and corrected, these reliability problems
may surface as security problems as well, as in the case of RowHammer,
especially if the technology is employed as part of the main memory
system.

We briefly examine two example potential vulnerabilities. We believe
future work examining these vulnerabilities, among others, are
promising for both fixing the vulnerabilities and enabling the
effective scaling of memory technology.

\subsection{Data Retention Failures}

Data retention is a fundamental reliability problem, and hence a
potential vulnerability, in charge-based memories like DRAM and flash
memory. This is because charge leaks out of the charge storage unit
(e.g., the DRAM capacitor or the NAND flash floating gate) over
time. As such memories become denser, three major trends make data
retention more difficult~\cite{raidr,dram-isca2013,kang-memforum2014,
  cai-hpca15}. First, the number of memory cells increases, leading to
the need for more refresh operations to maintain data
correctly. Second, the charge storage unit (e.g., the DRAM capacitor)
becomes smaller and/or morphs in structure, leading to potentially
lower retention times. Third, the voltage margins that separate one
data value from another become smaller (e.g., the same voltage window
gets divided into more ``states'' in NAND flash memory, to store more
bits per cell), and as a result the same amount of charge loss is more
likely to cause a bit error in a smaller technology node than a larger
one.

\subsubsection{DRAM Data Retention Issues}

Data retention issues in DRAM are a fundamental scaling limiter of the
DRAM technology~\cite{dram-isca2013, kang-memforum2014}.  We have
shown, in recent works based on rigorous experimental analyses of
modern DRAM
chips~\cite{dram-isca2013,samira-sigmetrics14,avatar-dsn15,khan-dsn16},
that determining the minimum retention time of a DRAM cell is getting
significantly more difficult. Thus, determining the correct rate at
which to refresh DRAM cells has become more difficult, as also
indicated by industry~\cite{kang-memforum2014}. This is due to two
major phenomena, both of which get worse (i.e., become more prominent)
with technology scaling. First, Data Pattern Dependence (DPD): the
retention time of a DRAM cell is heavily dependent on the data pattern
stored in itself and in the neighboring
cells~\cite{dram-isca2013}. Second, Variable Retention Time (VRT): the
retention time of some DRAM cells can change drastically over time,
due to a memoryless random process that results in very fast charge
loss via a phenomenon called trap-assisted gate-induced drain
leakage~\cite{yaney1987meta,restle1992dram, dram-isca2013}. These
phenomena greatly complicate the accurate determination of minimum
data retention time of DRAM cells. In fact, VRT, as far as we know, is
very difficult to test for because there seems to be no way of
determining that a cell exhibits VRT until that cell is observed to
exhibit VRT and the time scale of a cell exhibiting VRT does not seem
to be bounded, given the current experimental
data~\cite{dram-isca2013}. As a result, some retention errors can
easily slip into the field because of the difficulty of the retention
time testing.  Therefore, data retention in DRAM is a vulnerability
that can greatly affect both reliability and security of current and
future DRAM generations. We encourage future work to investigate this
area further, from both reliability and security, {\em as well as}
performance and energy efficiency perspectives. Various works in this
area provide insights about the retention time properties of modern
DRAM devices based on experimental
data~\cite{dram-isca2013,samira-sigmetrics14,avatar-dsn15,khan-dsn16,
  softmc}, develop infrastructures to obtain valuable experimental
data~\cite{softmc}, and provide potential solutions to the DRAM
retention time problem~\cite{raidr,
  dram-isca2013,samira-sigmetrics14,avatar-dsn15,khan-dsn16,
  memcon-cal16,darp-hpca2014}, all of which the future works can build
on.

Note that data retention failures in DRAM are likely to be
investigated heavily to ensure good performance and energy efficiency.
And, in fact they already are (see, for
example,~\cite{raidr,darp-hpca2014,memcon-cal16,samira-sigmetrics14,khan-dsn16}). We
believe it is important for such investigations to ensure no new
vulnerabilities (e.g., side channels) open up due to the solutions
developed.


\subsubsection{NAND Flash Data Retention Issues}

Experimental analysis of modern flash memory devices show that the
dominant source of errors in flash memory are data retention
errors~\cite{cai-date12}. As a flash cell wears out, its charge
retention capability degrades~\cite{cai-date12, cai-hpca15} and the
cell becomes leakier. As a result, to maintain the original data
stored in the cell, the cell needs to be refreshed~\cite{cai-iccd12,
  cai-itj2013}. The frequency of refresh increases as wearout of the
cell increases. We have shown that performing refresh in an adaptive
manner greatly improves the lifetime of modern MLC (multi-level cell)
NAND flash memory while causing little energy and performance
overheads~\cite{cai-iccd12, cai-itj2013}. Most high-end SSDs today
employ refresh mechanisms.

As flash memory scales to smaller nodes and even more bits per cell,
data retention becomes a bigger problem. As such, it is critical to
understand the issues with data retention in flash memory. Our recent
work provides detailed experimental analysis of data retention
behavior of MLC NAND flash memory~\cite{cai-hpca15}. We show, among
other things, that there is a wide variation in the leakiness of
different flash cells: some cells leak very fast, some cells leak very
slowly. This variation leads to new opportunities for correctly
recovering data from a flash device that has experienced an
uncorrectable error: by identifying which cells are fast-leaking and
which cells are slow-leaking, one can probabilistically estimate the
original values of the cells before the uncorrectable error
occurred. This mechanism, called {\em Retention Failure Recovery},
leads to significant reductions in bit error rate in modern MLC NAND
flash memory~\cite{cai-dsn15} and is thus very
promising. Unfortunately, it also points out to a potential security
and privacy vulnerability: by analyzing data and cell properties of a
failed device, one can potentially recover the original data. We
believe such vulnerabilities can become more common in the future and
therefore they need to be anticipated, investigated, and understood.

\subsection{Other Vulnerabilities in NAND Flash Memory}

We believe other sources of error (e.g., cell-to-cell interference)
and cell-to-cell variation in flash memory can also lead various
vulnerabilities. For example, another type of variation (that is
similar to the variation in cell leakiness that we described above)
exists in the vulnerability of flash memory cells to read
disturbance~\cite{cai-dsn15}: some cells are much more prone to read
disturb effects than others. This wide variation among cells enables
one to probabilistically estimate the original values of cells in
flash memory after an uncorrectable error has occurred. Similarly, one
can probabilistically correct the values of cells in a page by knowing
the values of cells in the neighboring
page~\cite{cai-sigmetrics14}. These mechanisms~\cite{cai-dsn15,
  cai-sigmetrics14} are devised to improve flash memory reliability
and lifetime, but the same phenomena that make them effective in doing
so can also lead to potential vulnerabilities, which we believe are
worthy of investigation to ensure security and privacy of data in
flash memories.

As an example, we have recently shown~\cite{cai-hpca17} that it is
theoretically possible to exploit vulnerabilities in flash memory
programming operations on existing solid-state drives (SSDs) to cause
(malicious) data corruption. This particular vulnerability is caused
by the {\em two-step programming} method employed in dense flash
memory devices, e.g., MLC NAND flash memory. An MLC device partitions
the threshold voltage range of a flash cell into four
distributions. In order to reduce the number of errors introduced
during programming of a cell, flash manufacturers adopt a two-step
programming method, where the least significant bit of the cell is
partially programmed first to some intermediate threshold voltage, and
the most significant bit is programmed later to bring the cell up to
its full threshold voltage.  We find that two-step programming exposes
new vulnerabilities, as both cell-to-cell program interference and
read disturbance can disrupt the intermediate value stored within a
multi-level cell before the second programming step completes. We show
that it is possible to exploit these vulnerabilities on existing
solid-state drives (SSDs) to alter the partially-programmed data,
causing (malicious) data corruption. We experimentally characterize
the extent of these vulnerabilities using contemporary 1X-nm (i.e.,
15-19nm) flash chips~\cite{cai-hpca17}. Building on our experimental
observations, we propose several new mechanisms for MLC NAND flash
that eliminate or mitigate disruptions to intermediate values,
removing or reducing the extent of the vulnerabilities, mitigating
potential exploits, and increasing flash lifetime by
16\%~\cite{cai-hpca17}. We believe investigation of such
vulnerabilities in flash memory will lead to more robust flash memory
devices in terms of both reliability and security, as well as
performance.

\section{Prevention}
\label{sec:prevention}

Various reliability problems experienced by scaled memory
technologies, if not carefully anticipated, accounted for, and
corrected, may surface as security problems as well, as in the case of
RowHammer.  We believe it is critical to develop principled methods to
understand, anticipate, and prevent such vulnerabilities. In
particular, principled methods are required for three major steps in
the design process.

First, it is critical to understand the potential failure mechanisms
and anticipate them beforehand. To this end, developing solid
methodologies for failure modeling and prediction is critical. To
develop such methodologies, it is essential to have real experimental
data from past and present devices. Data available both at the small
scale (i.e., data obtained via controlled testing of individual
devices, as in, e.g.,~\cite{dram-isca2013, aldram,
  samira-sigmetrics14, kevinchang-sigmetrics16, cai-date12,
  cai-date13, cai-dsn15, cai-hpca15, yixin-jsac16}) as well as at the
large scale (i.e., data obtained during in-the-field operation of the
devices, under likely-uncontrolled conditions, as in,
e.g.,~\cite{justin-memerrors-dsn15, justin-flash-sigmetrics15}) can
enable accurate models for failures, which could aid many purposes,
including the development of better reliability mechanisms and
prediction of problems before they occur.

Second, it is critical to develop principled architectural methods
that can avoid, tolerate, or prevent such failure mechanisms that can
lead to vulnerabilities. For this, we advocate co-architecting of the
system and the memory together, as we described earlier. Designing
intelligent, flexible, and configurable memory controllers that can
understand and correct existing and potential failure mechanisms can
greatly alleviate the impact of failure mechanisms on reliability,
security, performance, and energy efficiency. Described in
Section~\ref{sec:rowhammer-solutions}, this {\em system-memory
  co-design} approach can also enable new opportunities, like
performing effective processing near or in the memory
device~\cite{rowclone, vivek-and-or, tesseract, pei, gs-dram,
  tom-isca16, impica-iccd16, lazypim, lisa, pattnaik-pact16,
  emc-isca16, cre-micro16}. In addition to designing the memory device
together with the controller, we believe it is important to
investigate mechanisms for good partitioning of duties across the
various levels of transformation in computing, including system
software, compilers, and application software.

Third, it is critical to develop principled methods for electronic
design, automation and testing, which are in harmony with the failure
modeling/prediction and system reliability methods, which we mentioned
in the above two paragraphs. Design, automation and testing methods
need to provide high and predictable coverage of failures and work in
conjunction with architectural and across-stack mechanisms. For
example, enabling effective and low-cost {\em online profiling of
  DRAM}~\cite{dram-isca2013, samira-sigmetrics14, avatar-dsn15,
  khan-dsn16, memcon-cal16} in a principled manner requires
cooperation of failure modeling mechanisms, architectural methods, and
design, automation and testing methods.


\section{Conclusion}

It is clear that the reliability of memory technologies we greatly
depend on is reducing, as these technologies continue to scale to ever
smaller technology nodes in pursuit of higher densities. These
reliability problems, if not anticipated and corrected, can also open
up serious security vulnerabilities, which can be very difficult to
defend against, if they are discovered in the field. RowHammer is an
example, likely the first one, of a hardware failure mechanism that
causes a practical and widespread system security vulnerability. As
such, its implications on system security research are tremendous and
exciting. The need to prevent such vulnerabilities opens up new
avenues for principled approaches to 1) understanding, modeling, and
prediction of failures, and 2) architectural as well as design,
automation and testing methods for ensuring reliable operation. We
believe the future is very bright for research in reliable and secure
memory systems, and many discoveries abound in the exciting yet
complex intersection of reliability and security issues in such
systems.



\section*{Acknowledgments}

This paper, and the associated talk, are a result of the research done
together with many students and collaborators over the course of the
past 4-5 years. We acknowledge their contributions. In particular,
three PhD theses have shaped the understanding that led to this
work. These are Yoongu Kim's thesis entitled ``Architectural
Techniques to Enhance DRAM Scaling''~\cite{yoongu-thesis}, Yu
Cai's thesis entitled ``NAND Flash Memory: Characterization, Analysis,
Modeling and Mechanisms''~\cite{yucai-thesis} and his continued
follow-on work after his thesis, and Donghyuk Lee's thesis
entitled ``Reducing DRAM Latency at Low Cost by Exploiting
Heterogeneity''~\cite{donghyuk-thesis-arxiv16}. We also acknowledge
various funding agencies (NSF, SRC, ISTC, CyLab) and industrial
partners (AMD, Google, Facebook, HP Labs, Huawei, IBM, Intel,
Microsoft, Nvidia, Oracle, Qualcomm, Rambus, Samsung, Seagate, VMware)
who have supported the presented and other related work
generously over the years. The first version of this talk was
delivered at a CyLab Partners Conference in September 2015. Another
version of the talk was delivered as part of an Invited Session at DAC
2016, with a collaborative accompanying paper entitled ``Who Is the
Major Threat to Tomorrow’s Security? You, the Hardware
Designer''~\cite{dac-invited-paper16}.


\bibliographystyle{abbrv}
\small
\bibliography{conf}
\end{spacing}
\thispagestyle{plain}
\end{document}